# GaAs(111)A and B in hydrazine sulfide solutions : extreme polarity dependence of surface adsorption processes

V. L. Berkovits,[a] V. P. Ulin,[a] O. E. Tereshchenko,[b] D. Paget,[c+] A. C. H. Rowe,[c] P. Chiaradia,[d] B. P. Doyle[e] and S. Nannarone[e, f]

[a] A. F. Ioffe Physico-technical Institute, 194021 Saint Petersburg, Russia,

[b] Institute of Semiconductor Physics, Novosibirsk State University, 630090 Novosibirsk, Russia,

[c] Physique de la Matière Condensée, Ecole Polytechnique, CNRS, 91128 Palaiseau cedex, France,

[d] Dipartimento di Fisica, CNISM Unit and NAST, Università Tor Vergata, 00133 Roma, Italy,

[e] Laboratorio Nazionale TASC, INFM-CNR, s.s.14, km 163.5 in Area Science Park, 34012 Trieste, Italy,

[f] Dipartimento di Ingegneria dei Materiali e dell'Ambiente, Università di Modena e Reggio Emilia and CNISM, Via Vignolese 905, 41100 Modena, Italy,

Abstract: Chemical bonds formed by hydrazine-sulfide treatment of GaAs(111) were studied by synchrotron photoemission spectroscopy. At the B surface, the top arsenic atoms are replaced by nitrogen atoms, while GaAs(111)A is covered by sulfur, also bonded to underlying gallium, despite the sulfide molar concentration being $10^3$ times smaller than that of the hydrazine. This extreme dependence on surface polarity is explained by competitive adsorption processes of HS$^-$ and OH$^-$ anions and of hydrazine molecules, on Ga- adsorption sites, which have distinct configurations on the A and B surfaces.

*corresponding author, e-mail address*: daniel.paget@polytechnique.fr

The polar character of the (111) surfaces of III-V semiconductors is responsible for the distinct chemical behavior of the A and B surfaces[1] as manifested by differences in surface etching,[2] formation of Schottky barriers,[3] epitaxial growth,[4] electrochemical reactions,[5] and sulfide passivation.[6,7] In the present work, we show that polarity drastically affects the adsorption processes occurring during treatment of GaAs(111) by hydrazine sulfide solutions containing hydrazine molecules $N_2H_4$ (molar concentration M ~ 20), and extremely small amounts of $HS^-$ (M ~ $10^{-2}$) and $OH^-$ anions (M ~ $10^{-2}$). For GaAs(001), addition of small amounts of sulfide enables the removal of surface arsenic thereby providing adsorption of the hydrazine molecules on surface Ga atoms. Soft treatments by hydrazine sulfide solutions thus lead to formation of ultra-thin, chemically stable, GaN layers which electronically and chemically passivate the surface for up to several years,[8,9] and are a potential alternative to plasma nitridation.[10,11]

The chemical bonds established by the hydrazine-sulfide treatment on the (111)A and (111)B surfaces were analyzed using photoemission spectroscopy and low energy electron diffraction (LEED). Experiments were performed at the BEAR beamline of the ELETTRA storage ring (Trieste, Italy), and photoemission data were obtained using a hemispherical analyzer operated at constant pass energy. Two GaAs (111)A and (111)B surfaces were degreased and treated by $NH_4OH$ and HF in a glove box filled with nitrogen gas. A four hour chemical hydrazine-sulfide treatment was then performed after which the samples were dried in nitrogen, mounted next to each other on the same holder, and introduced into ultra-high vacuum without exposure to air. Measurements of core level spectra (CLS) were performed at 300 K for the as-treated surfaces and after annealing at 540 °C in order to remove weakly-bonded components, and to reveal the chemisorbed layers at the semiconductor surfaces. A very small oxygen concentration was initially found, which completely disappeared after annealing, and no carbon contamination was detected. Also studied for comparison were clean B and A GaAs(111) surfaces prepared by a HCl-isopropanol treatment.[12] The resulting Ga 3d (As 3d) CLS spectra coincided with those already published for Ga,[13] and As, [14] which allowed us to extract the bulk contributions to the CLS.

As seen in Fig. 1, a reacted component at higher binding energy, which disappears under annealing, is present in the As 3d CLS of the B surface and to some extent of the A one. Fig. 2, which shows the Ga3d CLS, exhibits for the two surfaces reacted components at higher binding energy. The S2p CLS of the A surface (inset in Fig. 3) is composed of a broad doublet, which only slightly changes under annealing. For the annealed B surface, the same CLS is reduced to a weak, well-defined doublet, while an intense doublet, of similar width, is also present on the as-treated surface at higher binding energy. The broader S2p CLS for the A surface is in agreement with the  fact that, as shown in Fig. 1, the 1x1 LEED pattern shown by the annealed A surface is less sharp than the one of the annealed B surface. Finally, the valence band (VB) spectra of the two annealed surfaces, together with those of clean surfaces of identical polarity, and their respective differences, are shown in Fig. 3. Again, the spectra are different for the two surface polarities.

As seen in the As 3d CLS in Fig. 1, after annealing at 540 °C the semiconductor surface is mostly gallium-terminated, *independently of the starting polarity*. Indeed the difference spectra Ab–bulk and Bb-bulk, obtained by subtracting the bulk contribution from the As 3d CLS of the two surfaces, do not exhibit reacted components at higher binding energy, but only a contribution at a positive chemical shift attributed to interstitial As. This which provides an avenue for  stress relaxation at the nitride/GaAs or sulfide/GaAs boundary.[15] The absence of reacted components in the As 3d CLS implies that As is removed from the B surface resulting in a Ga-rich surface upon which ions or molecules from the solution can be readily absorbed. This removal leaves unoccupied Ga-related lattice sites which are one-fold coordinated at the A surface and three-fold coordinated at the B one.

Although for the ultra-thin layers under consideration, direct observation of the N 1s CLS is not possible for sensitivity reasons,[16] the only candidate for chemical bonding to Ga on the B surface during hydrazine sulfide treatment is nitrogen.[8] Oxygen and carbon are absent. Sulfur predominantly lies in a relatively weakly bonded overlayer, and is responsible for the doublet at 158 eV binding energy in the S2p CLS, which disappears under annealing. After annealing, the concentration of sulfur atoms bonded to gallium is weak, as revealed from the small residual doublet at slightly lower binding energy. For the Ga 3d CLS, the difference spectrum Bb-bulk, obtained by subtracting the bulk contribution from the CLS after annealing, exhibits a surface component at a chemical shift of -0.7 eV (Fig. 2). From the ratio of the intensities of the chemically-shifted peaks in the Ga 3d CLS and of the bulk signal, and using an escape depth of 0.6 nm,[17] we estimate that the thickness of the adsorbed layer is 0.3 nm i.e. comparable with one monolayer. Observation of a sharp 1×1 LEED pattern (inset, Fig. 1) shows that the adsorbed layer is ordered and in registry with the substrate. The VB spectrum of the B surface after annealing at 540°C (Fig. 3), is similar to that measured after low doses of plasma-nitridization of GaAs(001) followed by annealing at 535°C.[11] The difference with the corresponding spectra of the clean surfaces exhibits a negative signal near 2 eV due to the presence at the clean surfaces of As dangling bonds.[13] The dominant contribution labeled $C_N$ at a binding energy of 5.1 eV is attributed to nitrogen. The contribution $C_S$ at a binding energy near 3.6 eV is attributed to sulfur,[18] that is in agreement with the smaller electronegativity of sulfur with respect to nitrogen. The weakness of the latter component is consistent with the weak residual S2p CLS signal.

The most striking results are obtained for the A surface, for which the adsorbed atoms are identified as sulfur, despite the fact that their concentration in solution is $10^3$ times smaller than that of hydrazine molecules. The weak annealing-induced change of the S2p CLS (Fig. 3) implies that most sulfur atoms are strongly bonded with the surface. The chemical shift of -0.5 eV for the reacted component in the Ga3d CLS (Fig. 2) is in qualitative agreement with the value of -0.4 eV observed for adsorption of sulfur atoms on clean (111)A GaAs.[19] The estimated thickness of the adsorbed layer is of 0.2 nm, i.e. comparable with one monolayer. The VB spectrum is dominated by a sulfur-related[19] peak near 3.5 eV. The difference with the spectrum of the clean surface, in addition to the negative As-related signal mostly exhibits the sulfur-related $C_s$ contribution, with a residual nitrogen-related $C_N$ one. The ratio of the two components suggests that the concentration of sulfur is larger than that of nitrogen by a factor of 3.

The extreme sensitivity of adsorption processes to the initial polarity of the (111) surface is a direct consequence of the fact that the reactions occur in a liquid electrolyte environment and thus strongly differ from those taking place in vacuum. In vacuum, nitride layers are formed on the A surface in the same way as on the B and (001) surfaces.[20] Nitrogen also forms bonds at Si(111) surfaces at room temperature.[21] In electrolytes, as a result of a charge exchange with the semiconductor,[22] there form empty Ga-related and, in contrast with vacuum adsorption, empty As-related dangling bonds. These bonds enable donor acceptor covalent bonding at surface adsorption sites denominated $V^{+3}_{Ga}$ and $V^{+3}_{As}$ in the case where three dangling bonds are present. Further, due to solvation of anions and molecules in solutions, adsorption processes from a liquid are characterized by a significantly lower change of potential energy than in vacuum. Thus, there occurs an intense particle exchange between the adsorbed layer and the solution, which here can involve several competitive, reversible and irreversible, processes, originating from $HS^-$ and $OH^-$ anions or from $N_2H_4$ molecules. These processes are described by the following reactions:

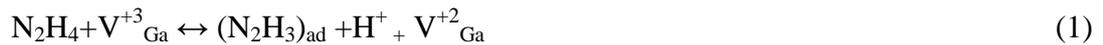
$$N_2H_4 + V^{+3}_{Ga} \leftrightarrow (N_2H_3)_{ad} + H^+ + V^{+2}_{Ga} \qquad\qquad (1)$$

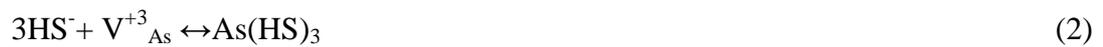
$$3HS^- + V^{+3}_{As} \leftrightarrow As(HS)_3 \qquad\qquad (2)$$

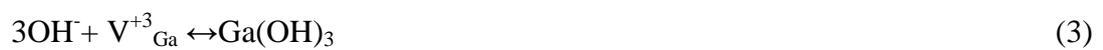
$$3OH^- + V^{+3}_{Ga} \leftrightarrow Ga(OH)_3 \qquad\qquad (3)$$

The present experimental results, obtained after a long duration treatment, should be viewed as the final state of a dynamic exchange depending on the relative efficiencies of the above surface adsorption-desorption processes.

Reaction (2) occurs in the same way as for treatment of GaAs(001) by sulfide solutions [22] and is shown from the chemical composition of weakly-bonded overlayers.[8] The annealing-induced changes of the As 3d CLS shown in the difference spectrum Ba-Bb of Fig. 1 exhibit a peak near 42.5 eV (chemical shift ~ -0.8 eV) characteristic of surface arsenic sulfides [22] and a peak at 44.5 eV (chemical shift ~ -2.8 eV) also attributed to As-S chemical bonds since its chemical shift is close to that found for $As_2S_3$.[23] At the A surface, a similar reaction, also observed to a minor extent,(see difference Aa-Ab in Fig. 1) concerns As atoms situated in the oxide layer.  In agreement with these findings the S2p CLS of the as-treated B surface exhibits a dominant doublet, attributed to As-S-H bonds, which lies at an energy 0.5 eV smaller than that of the residual S-Ga related doublet, also observed after annealing.[6]

Reaction (3) is confirmed by the presence of gallium oxide in the overlayer, as revealed from the Ga3d CLS (Fig. 2). The difference spectrum, Ba-Bb in Fig. 2 exhibits a reacted component near 20.6 eV which is oxygen-related.[24] The corresponding spectrum of the as-treated A surface exhibits a signal at 20.2 eV energy, also attributed to Ga-OH bonds since Ga atoms at the A surface only have one dangling bond available for $OH^-$ adsorption. This shows that $OH^-$ ions from the solution preferentially form bonds with surface gallium which can be subsequently desorbed, thus resulting in surface micro-etching.[9] Adsorption of $OH^-$ anions is in competition with the bonding of hydrazine molecules with surface Ga atoms. In the same way as for $OH^-$ adsorption, formation of a single Ga-N bond, according to Reaction (1) is reversible. However, in contrast with $OH^-$, nitrogen atoms of hydrazine molecules can form a second and a third chemical bond. Such a process is only possible on the B surface where nitrogen replaces surface As and can therefore establish three bonds with underlying gallium atoms, thus forming a monolayer of nitride irreversibly bonded to the surface.

On the A surface, since Ga atoms have one dangling bond perpendicular to the surface, irreversible adsorption of nitrogen or sulfur atoms from electrolytes through multiple bond formation is not energetically possible.[10] Formation of doubly-bonded dimers (Ga-X-X-Ga) should lead to a more stable configuration. Sulfur dimers are more likely to form than N dimers because both X-X and Ga-X distances, to be compared with the distance of 0.4 nm between surface Ga, are larger for S (0.21 and 0.22 nm) than for N (0.14 and 0.19 nm) and because the X-X formation energy is larger for S (2.8 eV) than for N (0.8 eV). Further, chains of sulfur atoms,$(S_n)$ known to be present in the solution, can also be bonded to the surface, leading to Ga-S-…-S-Ga configurations. The bonded sulfur chains or dimers are randomly oriented thereby leading to the more diffuse LEED pattern and the increased width of the S2p signal.

It is concluded that the extreme selectivity as a function of polarity for adsorption of nitrogen at (111)GaAs is caused by competitive adsorption of $OH^-$ anions and hydrazine molecules on surface Ga. Note finally that formation of one complete adsorbed layer of group V elements on the A surface at top sites at equilibrium would appear thermodynamically improbable. Such a formation would break the fundamental rule according to which, in the (111) direction, thermodynamically stable zinc-blende crystals are composed of biatomic layers bonded along the (111) direction by cation-anion bonds.

**Acknowledgments:**

The work was partly supported by the Program of the Presidium of the Russian Academy of Sciences, by the French Embassy in Moscow (V.L.B. and O.E.T) and by the Agence Nationale de la Recherche (06-BLAN-0253-03).

## Figures

Fig. 1: As 3d core level signals of as-treated GaAs(111) (Xa, where X stands for A or B) and after annealing to 540°C (Xb) taken at a photon energy of 100 eV. The bulk signal is obtained from the CLS of the clean (111)A surface, measured independently. The differences show the effect of annealing and the reacted components at the annealed surfaces. Also shown in the inserts are the LEED patterns after annealing.

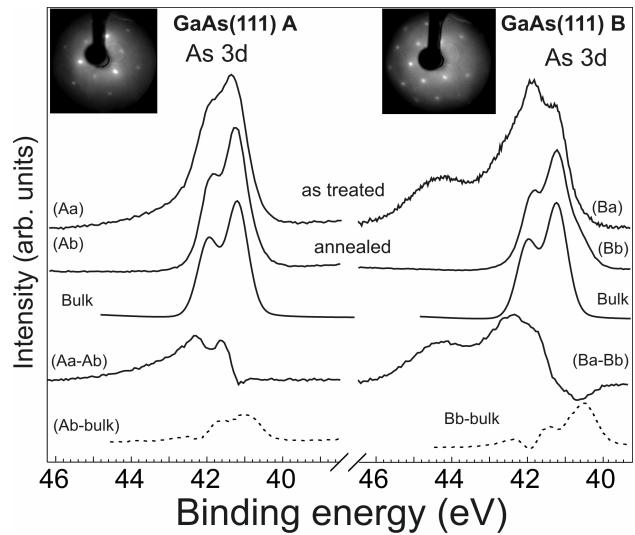

Fig. 2 : The top spectra show the Ga 3d core level signals of as-treated GaAs(111) (Xa) and after annealing to 540°C (Xb) taken at a photon energy of 100 eV. The difference spectra reveal the reacted components for the two surfaces. The center graphs compare the CLS of the annealed surfaces with the bulk contribution.

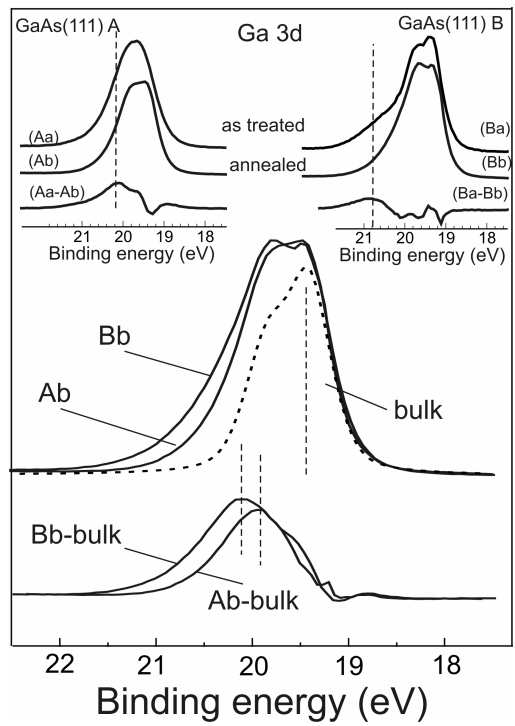

Fig. 3: VB spectra of the annealed (111)A and B surfaces, together with corresponding spectra for clean surfaces and their differences showing the S and N contributions.(photon energy 100 eV) The insets show the S 2p core level signals of as-treated (Xa) and annealed surfaces (Xb), where X (A or B) is the surface polarity and their deconvolutions. The photon energy is 250 eV.

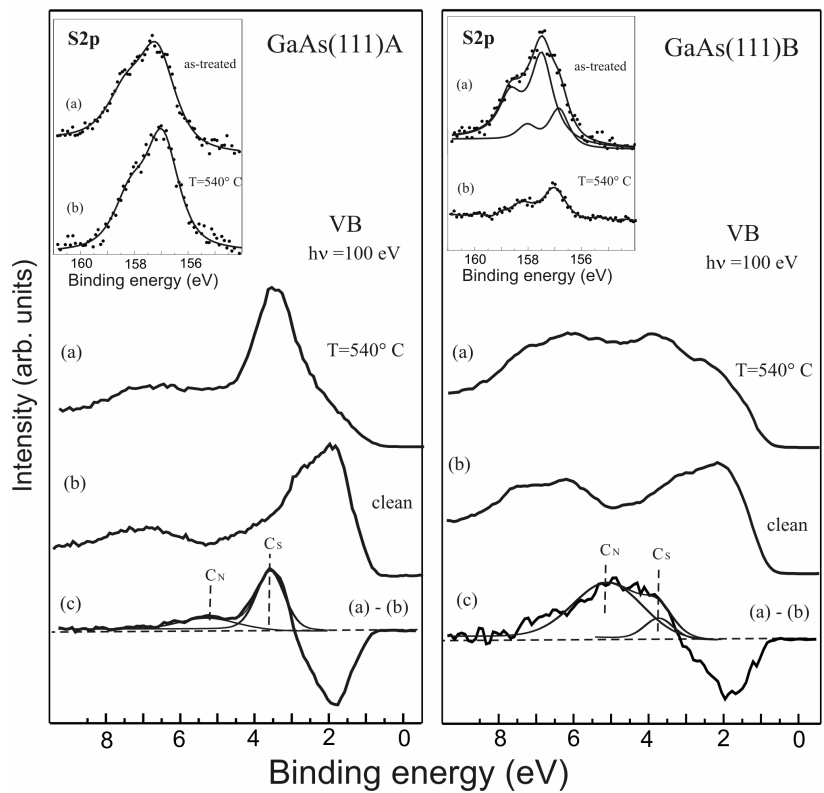